\documentclass[conference]{IEEEtran}
\IEEEoverridecommandlockouts
\usepackage[utf8]{inputenc}
\usepackage{cite,balance}
\usepackage{amsmath,amssymb,amsfonts,multirow,upgreek,subcaption}
\usepackage[table]{xcolor}
\usepackage{textcomp}
\usepackage{xcolor}
\usepackage{bm}
\usepackage{algpseudocode}
\usepackage{algorithm}
\usepackage{enumitem}
\usepackage{psfrag}
\usepackage{tikz}
\usetikzlibrary{shapes.geometric, arrows}
\tikzstyle{roundrec} = [font=\footnotesize,rectangle, rounded corners, minimum width=1cm, minimum height=1cm,text centered, text width=2.5cm, draw=black]
\tikzstyle{rec} = [font=\footnotesize,rectangle, minimum width=3cm, minimum height=1cm, text width=3.75cm, draw=black]
\tikzstyle{bubble} = [font=\footnotesize,rectangle, rounded corners, text centered, draw=black]
\tikzstyle{arrow} = [thick,->,>=stealth]
\def\BibTeX{{\rm B\kern-.05em{\sc i\kern-.025em b}\kern-.08em
    T\kern-.1667em\lower.7ex\hbox{E}\kern-.125emX}}

\renewcommand{\j}{\mathrm{j}}
\DeclareMathOperator*{\argmax}{arg\,max}

\newcommand*{\herm}{^{\mathsf{H}}}
\newcommand*{\transp}{^{\mathsf{T}}}

\title{Adaptive Downlink Localization in \\ Near-Field and Far-Field}

\author{\IEEEauthorblockN{
Georgios Mylonopoulos\IEEEauthorrefmark{1}\IEEEauthorrefmark{2},   
Behrooz Makki\IEEEauthorrefmark{2},   
Gábor Fodor\IEEEauthorrefmark{3},   
Stefano Buzzi\IEEEauthorrefmark{1}\IEEEauthorrefmark{4},    
}                                  

\IEEEauthorblockA{\IEEEauthorrefmark{1}
University of Cassino and Southern Latium, Cassino, Italy \& CNIT, Parma, Italy}
\IEEEauthorblockA{\IEEEauthorrefmark{2}
Ericsson Research, Ericsson AB,  Göteborg}
\IEEEauthorblockA{\IEEEauthorrefmark{3}
Ericsson Research, Ericsson AB,  Stockholm, Sweden \& KTH Royal Institute of Technology, Stockholm, Sweden}
\IEEEauthorblockA{\IEEEauthorrefmark{4}
Politecnico di Milano, Milan, Italy\\ 
E-mail: \{georgios.mylonopoulos, buzzi\}@unicas.it, \{behrooz.makki, gabor.fodor\}@ericsson.com}
}

\begin{document}

\maketitle

\begin{abstract}
This paper considers the problem of downlink localization of user equipment devices (UEs) that are either in the near-field (NF) or in the far-field (FF) of the array of the serving base station (BS). We propose a dual signaling scheme, which can be implemented at the BS, for localizing such UEs. The first scheme assumes FF, while the other assumes NF conditions. Both schemes comprise a beam-sweeping technique, employed by the BS, and a localization algorithm, employed by the UEs. The FF-based scheme enables beam-steering with a low signaling overhead, which is utilized for the proposed localization algorithm, while the NF-based scheme operates with a higher complexity. Specifically, our proposed localization scheme takes advantage of the relaxed structure of the FF, which yields low computational complexity, but is not suitable for operating in the NF. Since the compatibility and the performance of the FF-based scheme depends on the BS-to-UE distance, we study the limitations of FF-based procedure, explore the trade-off in terms of performance and resource requirements for the two schemes, and propose a triggering condition for operating the component schemes of the dual scheme. Also, we study the performance of an iterative localization algorithm
that takes into account the accuracy-complexity trade-off and adapts to the actual position of the UE. We find that the conventional Fraunhofer distance is not sufficient for adapting localization algorithms in the mixed NF and FF environment.
\end{abstract}

\section{Introduction}

Evolving 6G networks are expected to provide high accuracy positioning and sensing services, with several opportunities and challenges~\cite{de2021convergent}. Extracting positional information is strongly tied to our ability to accurately model the propagation environment~\cite{de2021convergent}. In terms of localization, the far-field (FF) and near-field (NF) channel model mismatch is analyzed in~\cite{chen2022channel}. FF conditions are validly expected in 5G networks, but larger arrays and higher frequency bands in 6G networks will lead to user equipment devices (UEs) being in the base station’s (BS) NF region~\cite{cui2022near}. Localization in the NF has been explored, with several techniques employed to account for the NF channel model~\cite{jingjing2021search,su2021deep,shu2020near}. FF conditions provide a simpler channel model and, hence, lower complexity positioning algorithms. Under the FF assumption, downlink (DL) localization with a Cram\'er-Rao analysis is explored in~\cite{fascista2021downlink}.

As the NF region increases when large antenna arrays are deployed, a mixed-model system needs to be explored. When the communication problem is considered, and the UE approaches the large BS array, it is essential to utilize accurate channel models~\cite{de2020near}. 
In~\cite{he2021mixed,huang2020one}, radio source localization has been explored in a mixed FF and NF model. Positioning accuracy is also improved from the true channel model. NF beam-focusing and traditional FF beam-steering techniques are inherently different and there are several beam-forming challenges that need to be addressed~\cite{zhang20236g}. Recognizing these aspects, the FF beamforming in the NF region and the limited array gain relative to the Fraunhofer distance (FD) have been recently explored in~\cite{bjornson2021primer}. Joint FF and NF localization needs to be investigated, exploring the degrees of freedom and the limitations of the simplified FF channel model.

In this paper, we develop two distinctive signaling schemes and localization algorithms, accounting for FF or NF conditions. Assuming a large BS-to-UE distance we introduce the FF-based scheme. We exploit the relaxed FF channel model and utilize a beam-steering technique, resulting in lower signaling overhead and computational complexity. In contrast, our NF-based scheme always considers the true channel model and extensively covers the whole region utilizing beam-focusing technique. With this intuition, we propose an adaptive signaling scheme and pose an accuracy-complexity trade-off. We study the range in which the NF- and FF-based schemes can be successfully employed, and compare it with the traditional FD boundary. We analyse the complexity of our proposed schemes and evaluate their performance. Our results show that our proposed adaptive localization scheme balances the positioning accuracy and resource requirements of the system. Moreover, the distance where the NF-based scheme is \emph{necessary} diverges from the FD and depends, among other factors, on the considered geometry and beam-sweeping techniques of each signaling scheme.

\begin{figure}[h]
    \centering
    \includegraphics[width=\columnwidth]{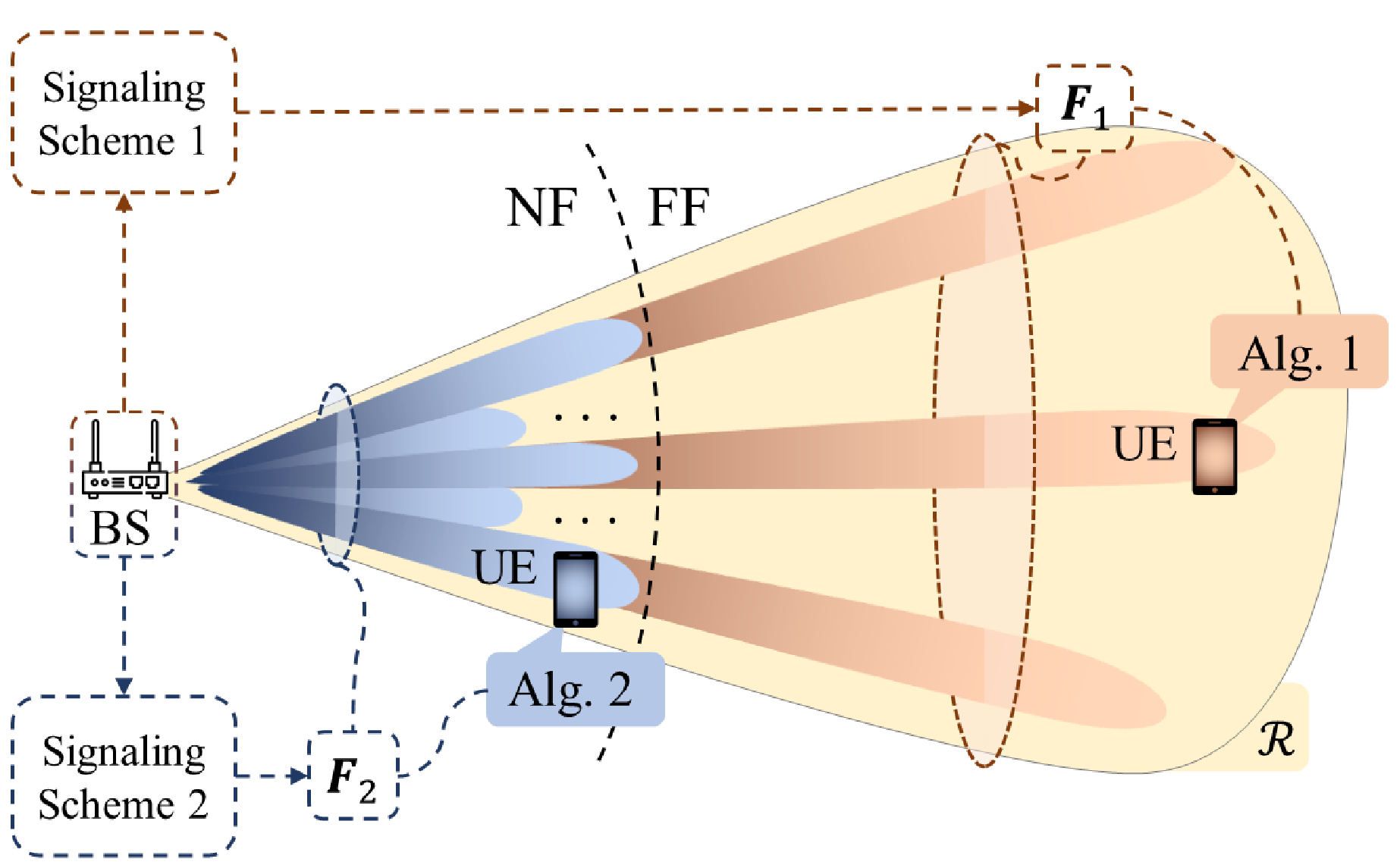}
    \caption{Considered scenario where BS has 2 signaling options, designed for DL UE localization in the FF and the NF.}
    \label{fig:sm}
\end{figure}

\section{System Model}
In this section, we describe the considered system and introduce the two channel models and beam-sweeping strategies that correspond to the FF or NF assumptions. Figure~\ref{fig:sm} shows the geometrical setup. There is a BS at point $\bm{p}_{\rm{BS}}$ equipped with a uniform planar array of $N_{\rm{BS}} = N_{\rm{BS}}^{(\rm{x})} \times N_{\rm{BS}}^{(\rm{z})}$ antennas, with an inter-antenna spacing of $\delta_{\rm{BS}}$, serving single antenna UEs within the region $\mathcal{R}$. The UEs estimate their own position, $\bm{p}_{\rm{UE}}$ and forward it to the BS, where the localization procedure is coordinated. 

\begin{figure}
    \centering
    \begin{tikzpicture}[node distance=2cm]
        \node (start) at (1.5,9.5) [roundrec] {The BS employs one \emph{Signaling Scheme}};
        \node (scheme1) at (-0.9,8.25) [rec] {\textbf{FF-based signaling}:\\\textbullet The BS uses $\bm{F}_{1}$.\\\textbullet The UEs use Alg. 1 for $\hat{\bm{p}}_{\rm{UE}}$.};
        \node (scheme2) at (3.85,8.25) [rec] {\textbf{NF-based signaling}:\\\textbullet The BS uses $\bm{F}_{2}$.\\\textbullet The UEs use Alg. 2 for $\hat{\bm{p}}_{\rm{UE}}$.};
        \node (dec) at (-0.9,7.1) [bubble] {$\hat{d}_{\rm{UE}} > d_{\rm{NF}}$};
        \node (end) at (1.5,5.75) [bubble] {Process completed};
        \node (init) at (1.5,10.5) [bubble] {Initiate Process};
        
        \draw [arrow] (node cs:name=init,anchor=south) -- (node cs:name=start,anchor=north);
        \draw [arrow] (node cs:name=start,anchor=west) --(-0.9,9.5)-- node[xshift=-0.5cm,yshift=0.6cm] {\footnotesize BS opts for scheme 1}(node cs:name=scheme1,anchor=north);
        \draw [arrow] (node cs:name=start,anchor=east) --(3.85,9.5)-- node[xshift=0.5cm,yshift=0.6cm] {\footnotesize BS opts for scheme 2}(node cs:name=scheme2,anchor=north);
        \draw [arrow] (node cs:name=scheme1,anchor=south) -- (node cs:name=dec,anchor=north);
        \draw [arrow] (node cs:name=dec,anchor=west) --node[anchor=north] {\footnotesize True}(-2.75,7.1)--(-2.75,6.35)--(1.5,6.35)-- (node cs:name=end,anchor=north);
        \draw [arrow] (node cs:name=dec,anchor=east) --node[anchor=north] {\footnotesize False}(1.5,7.1)--(1.5,8.25)-- (node cs:name=scheme2,anchor=west);
        \draw [arrow] (node cs:name=scheme2,anchor=south) --(3.85,6.35)--(1.5,6.35)-- (node cs:name=end,anchor=north);
    \end{tikzpicture}
    \caption{Proposed DL localization signaling procedure. The estimated DOF is $\hat{d}_{\rm{UE}}$ and the distance where the NF-based scheme has to be employed is denoted by $d_{\rm{NF}}$.}
    \label{fig:loc_proc}
\end{figure}
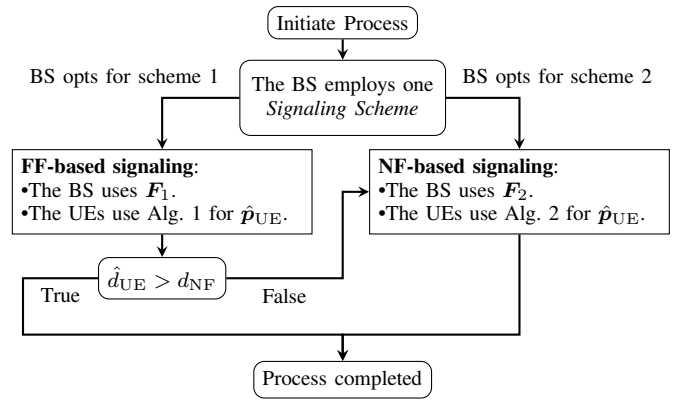

The localization procedure is summarised in Fig.~\ref{fig:loc_proc}, with two distinctive signaling schemes available to the BS:
\begin{itemize}
\item \emph{FF-based signaling}: The BS transmits $J_{1}$ OFDM pilots, employing the beam-steering matrix $\bm{F}_{1} \in \mathbb{C}^{N_{\rm{BS}}\times J_{1}}$. The design of $\bm{F}_{1}$ assumes FF conditions.
\item \emph{NF-based signaling}: The BS transmits $J_{2}$ OFDM pilots, employing the beam-focusing matrix $\bm{F}_{2} \in \mathbb{C}^{N_{\rm{BS}}\times J_{2}}$. The design of $\bm{F}_{2}$ assumes NF conditions.
\end{itemize}
The OFDM pilots consist of $Q$ subcarriers. The subcarrier bandwidth (BW) is $W_{\rm{o}}$ with a subcarrier spacing of $W_{\rm{sub}}$.

\subsection{Channel \& Signal Model}
Here, we introduce the true and the relaxed channel models. The true -NF- channel between the BS and the $m^{\rm{th}}$ UE for the $q^{\rm{th}}$ subcarrier is given by
\begin{subequations}
\begin{align}
\bm{H}_{\rm{NF}}^{(m)} &= [\bm{h}_{1}^{(m)} \dotsc \bm{h}_{q}^{(m)} \dotsc \bm{h}_{Q}^{(m)}] \in \mathbb{C}^{N_{\rm{BS}}\times Q} , \\
\bm{h}_{q}^{(m)} &= \beta^{(m)} \bm{\alpha}_{q}(\bm{p}_{\rm{UE}}^{(m)}) \in \mathbb{C}^{N_{\rm{BS}}\times 1} ,
\end{align}
\label{eq:channel_h_NF}
\end{subequations}
where $\bm{\alpha}_{q}(\bm{p})$ is the steering vector for point $\bm{p}$, relative to the BS and $\beta^{(m)}$ is a scalar accounting for the free-space path loss and the synchronization error for the $m^{\rm{th}}$ UE, given by
\begin{align}
\beta^{(m)} =  \frac{\lambda_{\rm{o}}}{4\pi d^{(m)}} \text{exp}\bigl\{-\j \phi_{\rm{o}}^{(m)}\bigr\}.
\label{eq:beta_qm}
\end{align}
Here, $\lambda_{\rm{o}}$ is the wavelength, $d^{(m)} = \| \bm{p}_{\rm{BS}} - \bm{p}_{\rm{UE}}^{(m)}\|$ is the distance of flight (DOF) for the $m^{\rm{th}}$ UE and $\phi_{\rm{o}}$ a uniformly random phase offset, accounting for the synchronization error. The $n^{\rm{th}}$ element of the steering vector, $\bm{\alpha}_{q}(\bm{p})$, is  
\begin{align}
\bigl[ \bm{\alpha}_{q}(\bm{p})\bigr]_{n} = \frac{\lambda_{q}d^{(m)}}{\lambda_{\rm{o}}\|\bm{p} - \bm{p}_{\rm{BS}}^{(n)} \|}\text{exp}\bigl\{ -\j\frac{2\pi}{\lambda_{q}} \|\bm{p} - \bm{p}_{\rm{BS}}^{(n)} \|\bigr\} \quad , 
\label{eq:sv_NF}
\end{align}
where $\bm{p}_{\rm{BS}}^{(n)}$ is the position of the $n^{\rm{th}}$ BS antenna, for $n = 1,\dots,N_{\rm{BS}}$. The channel in~\eqref{eq:channel_h_NF} preserves the spherical nature of the wavefront and requires no assumptions. If we assume a sufficiently large distance, an approximation  is given by 
\begin{subequations}
\begin{align}
\tilde{\bm{h}}_{q}^{(m)} &= \beta^{(m)}\text{exp}\bigl\{-\j2\pi(q-1)\frac{W_{\rm{sub}}}{c_{\rm{o}}}d^{(m)}\bigr\}\tilde{\bm{\alpha}}(\bm{\theta}_{\rm{UE}}^{(m)}), \\
\bm{H}_{\rm{FF}}^{(m)} &= [\tilde{\bm{h}}_{1}^{(m)} \dotsc \tilde{\bm{h}}_{q}^{(m)} \dotsc \tilde{\bm{h}}_{Q}^{(m)}] \quad , \notag \\
&= \beta^{(m)}\bm{t}\transp(d^{(m)})\tilde{\bm{\alpha}}(\bm{\theta}_{\rm{UE}}^{(m)}) \in \mathbb{C}^{N_{\rm{BS}}\times Q} ,
\end{align}
\label{eq:channel_h_FF}
\end{subequations}
where $\tilde{\bm{\alpha}}(\bm{\theta})$ is the planar wave steering vector for an angle-of-departure (AOD), $\bm{\theta} = [\theta^{\rm{az}}, \theta^{\rm{el}}]$ and 
\begin{align}
\bigl[ \bm{t}(d)\bigr]_{q} = \text{exp}\bigl\{-\j2\pi(q-1)\frac{W_{\rm{sub}}}{c_{\rm{o}}}d\bigr\} \quad .
\label{eq:t}
\end{align}
The relaxed model in~\eqref{eq:channel_h_FF} is the FF model and the AOD and the DOF are decoupled.
The received signal for the $j^{\rm{th}}$ spatial stream along the $q^{\rm{th}}$ subcarrier is given by
\begin{align}
y_{q,j}^{(m)} = \sqrt{\mathcal{P}} \bm{f}_{j}\herm\bm{h}_{q}^{(m)}x_{q,j}^{(m)} + z_{q,j} \quad ,
\label{eq:y_qj}
\end{align}
where $\mathcal{P}$ is the transmit power, $\bm{f}_{j}$ is the $j^{\rm{th}}$ precoding vector of the employed signaling scheme, $x_{q,j}^{(m)}$ is the OFDM symbol and $z_{q,j} \sim \mathcal{CN}(0,\sigma_{z}^2)$ denotes noise. The received signal can be organized in a matrix form, $\bm{Y}\in\mathbb{C}^{J\times Q}$, while the vectors $\bm{y}_{j}\in\mathbb{C}^{J\times 1}$ and $\bm{y}_{q}\in\mathbb{C}^{Q\times 1}$ are the collections of scalars for the $j^{\rm{th}}$ beam and the $q^{\rm{th}}$ subcarrier, respectively.

\section{UE Localization}
In this section, we introduce the two localization algorithms, we explore their complexity and investigate their practicality. Compared to the NF channel model, the FF model allows for a localization algorithm of lower complexity, at the cost of accuracy. The lower signaling overhead also affects the computational load. With our proposed scheme, the high complexity algorithm is used \emph{only when necessary}.

\subsection{Localization Algorithms}
Algorithm~\ref{alg:pos_scheme_1} explores the FF-based localization procedure. A weighting procedure across the $J_{1}$ beams, with the normalized weight of each beam, $w_{j}$, acts as an RSSI, encapsulating how much energy is received from each beam of $\bm{F}_{1}$. For a normalised RSSI below a threshold, $\varepsilon_{w}$, we disregard the corresponding beam, as it may lead to unreliable estimations. For the position estimation we introduce an iterative estimation procedure along the DOF and AOD. The relaxed channel model in~\eqref{eq:channel_h_FF} decouples the range and the angular domain. The DOF estimation relies on the phase offset between subcarriers and each beam is processed individually, resulting into $J_{1}$ DOF estimations. The RSSI indicator provides a weighted mean for each iteration. On the other hand, the AOD estimation relies on the phase offset between antenna elements and each subcarrier is processed individually. The computational complexity depends on the number of iterations $I_{1}$, the number of search steps in each maximization process, $I_{d}$ and $I_{\theta}$, the number of spatial beams, $J_{1}$, and the computational cost of reconstructing a FF steering vector and a subcarrier phase offset vector in~\eqref{eq:t}. See Section~\ref{sec:complexity} for a detailed discussion and comparison with the NF localization algorithm.  

\begin{algorithm}
\caption{Localization under the FF assumption}\label{alg:pos_scheme_1} 
\begin{algorithmic}
\State Define $\varepsilon_{w} \in\mathbb{R}, I_{1} , I_{d} , I_{\theta} \in\mathbb{N}$
\State Input: $\bm{Y}, \bm{p}_{\rm{BS}}, \mathcal{R}$ 
\For{$j=1:J_{1}$}
    \State $w_{j} = \|\bm{y}_{j}\|/\text{max}(\|\bm{y}_{1:J_{1}}\|)$
    \If{$w_{j} < \varepsilon_{w}$}
        \State $w_{j} = 0$
    \EndIf
\EndFor
\State $\tilde{\bm{F}}_{1} = \bm{F}_{1}\rm{diag}(\bm{w})$
\For{$ii = 1:I_{1}$}
    \For{$j=1:J_{1}$} \Comment{$I_{d}$ search steps}
        \State $\hat{d}_{j} = \argmax_{d} \| \bm{y}\herm_{j}\bm{t}(d)\|$ 
    \EndFor
    \State $\hat{d} = \sum_{j=1}^{J_{1}}w_{j}\hat{d}_{j} / \sum_{j=1}^{J_{1}}w_{j}$
    \For{$q=1:Q$} \Comment{$I_{\theta}$ search steps}
        \State $\hat{\theta}^{\rm{az}}_{q} = \argmax_{\theta^{\rm{az}}} \| \bm{\alpha}(\bm{\theta})\herm\tilde{\bm{F}}_{1}\bm{y}_{q} \| : \bm{\theta}=[\theta^{\rm{az}}, \hat{\theta}^{\rm{el}}]$ 
    \EndFor
    \State $\hat{\theta}^{\rm{az}} = \sum_{q=1}^{Q} \hat{\theta}^{\rm{az}}_{q}/ Q$
    \For{$q=1:Q$} \Comment{$I_{\theta}$ search steps}
        \State $\hat{\theta}^{\rm{el}}_{q} = \argmax_{\theta^{\rm{el}}} \| \bm{\alpha}(\bm{\theta})\herm\tilde{\bm{F}}_{1}\bm{y}_{q} \| : \bm{\theta}=[\hat{\theta}^{\rm{az}}, \theta^{\rm{el}}]$ 
    \EndFor
    \State $\hat{\theta}^{\rm{az}} = \sum_{q=1}^{Q} \hat{\theta}^{\rm{az}}_{q}/ Q$

\EndFor
\State $\hat{\bm{p}}_{\rm{UE}} = \bm{p}_{\rm{BS}} + \hat{d} \begin{bmatrix}
\cos(\hat{\theta}^{\rm{el}})\sin(\hat{\theta}^{\rm{az}})\\ 
\cos(\hat{\theta}^{\rm{el}})\cos(\hat{\theta}^{\rm{az}})\\ 
\sin(\hat{\theta}^{\rm{el}})
\end{bmatrix}$ \Comment{Output}
\end{algorithmic}
\end{algorithm}

Algorithm~\ref{alg:pos_scheme_2} highlights the NF-based localization procedure. An iterative maximization along the DOF and AOD provides the position estimation. We choose the focus point of the spatial stream of $\bm{F}_{2}$ with the highest RSSI as a starting point. Similar to Alg.~\ref{alg:pos_scheme_1}, the beams with an RSSI below a thresehold are disregarded. The focus-points of $\bm{F}_{2}$ need to be dense enough to ensure proper coverage of $\mathcal{R}$, which requires a larger $J_{2}$ and thus a higher signaling overhead. The computational complexity is affected by the increased signaling overhead. In Alg.~\ref{alg:pos_scheme_2} each maximization search is performed across each subcarrier separately. Note that each search step in Alg.~\ref{alg:pos_scheme_2} requires the reconstruction of the NF steering vector, $\bm{\alpha(\bm{p})}$, for both the angular and the range domain. 

\begin{algorithm}
\caption{Localization under the NF assumption}\label{alg:pos_scheme_2} 
\begin{algorithmic}
\State Define $\varepsilon_{w} \in\mathbb{R}, I_{2} , I_{d} , I_{\theta} \in\mathbb{N}$
\State Input: $\bm{Y}, \bm{p}_{\rm{BS}}, \mathcal{R}$ 
\For{$j=1:J_{2}$}
    \State $w_{j} = \|\bm{y}_{j}\|/\text{max}(\|\bm{y}_{1:J_{2}}\|)$
    \If{$w_{j} < \varepsilon_{w}$}
        \State $w_{j} = 0$
    \EndIf
\EndFor
\State $j\ast = \argmax_{j} w_{j} $
\State $\hat{\bm{p}} = \bm{p}_{j\ast}$ \Comment{focus point of $\bm{f}_{2,j\ast}$}
\State $\tilde{\bm{F}}_{2} = \bm{F}_{2}\rm{diag}(\bm{w})$
\For{$ii = 1:I_{2}$}
    \For{$q=1:Q$} \Comment{$I_{d}$ search steps}
        \State $\hat{d}_{q} = \argmax_{d} \| \bm{\alpha}(\bm{p})\herm\tilde{\bm{F}}_{2}\bm{y}_{q} \|$ 
    \EndFor
    \State $\hat{d} = \sum_{q=1}^{Q}\hat{d}_{q} / Q$
    \For{$q=1:Q$} \Comment{$I_{\theta}$ search steps}
        \State $\hat{\theta}^{\rm{az}}_{q} = \argmax_{\theta^{\rm{az}}} \| \bm{\alpha}(\bm{p})\herm\tilde{\bm{F}}_{2}\bm{y}_{q} \|$ 
    \EndFor
    \State $\hat{\theta}^{\rm{az}} = \sum_{q=1}^{Q} \hat{\theta}^{\rm{az}}_{q}/ Q$
    \For{$q=1:Q$} \Comment{$I_{\theta}$ search steps}
        \State $\hat{\theta}^{\rm{el}}_{q} = \argmax_{\theta^{\rm{el}}} \| \bm{\alpha}(\bm{p})\herm\tilde{\bm{F}}_{2}\bm{y}_{q} \|$ 
    \EndFor
    \State $\hat{\theta}^{\rm{az}} = \sum_{q=1}^{Q} \hat{\theta}^{\rm{az}}_{q}/ Q$

\EndFor
\State $\hat{\bm{p}}_{\rm{UE}} = \bm{p}_{\rm{BS}} + \hat{d} \begin{bmatrix}
\cos(\hat{\theta}^{\rm{el}})\sin(\hat{\theta}^{\rm{az}})\\ 
\cos(\hat{\theta}^{\rm{el}})\cos(\hat{\theta}^{\rm{az}})\\ 
\sin(\hat{\theta}^{\rm{el}})
\end{bmatrix}$ \Comment{Output}
\end{algorithmic}
\end{algorithm}

\subsection{Complexity Analysis}\label{sec:complexity}
The alternating maximization and the 3-dimensional grid search is essentially a non-convex optimization problem and it poses the main computational load for both algorithms. The search resolution and the number of iterations affect the computational requirements of such a optimization procedure. The relevant computational requirements are introduced in Table~\ref{tab:complexity}. The size of the BS array and the number of transmitted beams also increase the overall complexity. Other than $J_{2}$ being larger than $J_{1}$, the distinctive disadvantage of Alg.~\ref{alg:pos_scheme_2} is the need to reconstruct the NF $\bm{\alpha}$, which is notably more complex than the FF $\tilde{\bm{\alpha}}$. 

\begin{table}
\caption{\rule{0mm}{7mm}Complexity analysis of Algorithms~\ref{alg:pos_scheme_1} \& \ref{alg:pos_scheme_2}.}
\label{tab:complexity}
\begin{center}
\renewcommand{\arraystretch}{1.5} 
\vspace{-0.3 cm}
\begin{tabular}{l l}
    \hline
    \hline	
    $\mathcal{O}\{ \bm{t}(d)\}$ & $\propto Q$ : One set of multiplications per subcarrier. \\
    \hline
    $\mathcal{O}\{ \tilde{\bm{\alpha}}(\bm{\theta})\}$ & $\propto N_{\rm{BS}}$ : One set of multiplications per antenna element.\\ 
    \hline
    $\mathcal{O}\{ \bm{\alpha}(\bm{p})\}$ & $\propto N_{\rm{BS}} \cdot\mathcal{O}\{ \|\bm{u} - \bm{v} \|\}$ : One set of multiplications and \\ &\quad an exact distance to be derived per antenna element.\\
    \hline
    $\mathcal{O}\{ \rm{Alg.~\ref{alg:pos_scheme_1}} \}$ & $\propto I_{1}I_{d}J_{1}\bigl(\mathcal{O}\{ \bm{t}(d)\} + Q\bigr) +$\\&\quad$ 2I_{1}I_{\theta}Q\bigl(\mathcal{O}\{ \tilde{\bm{\alpha}}(\bm{\theta})\} + N_{\rm{BS}}^{2}J_{1}\bigr)$ \\
    \hline
    $\mathcal{O}\{ \rm{Alg.~\ref{alg:pos_scheme_2}} \}$ & $\propto I_{1}Q(I_{d}+2I_{\theta})\bigl(\mathcal{O}\{ \bm{\alpha}(\bm{p})\}+ N_{\rm{BS}}^{2}J_{2}\bigr)$ \\
    \hline
\end{tabular}
\end{center}
\end{table}

\section{Simulation Results}\label{sec:sim}
In this section, we present simulation evaluations and different simulation setups that explore the performance of the two proposed schemes and highlight the complexity-accuracy trade-off.The simulation parameters given in Table~\ref{tab:sim_par}.
\begin{table}
\caption{Simulation Parameters}
\label{tab:sim_par}
\begin{center}
\vspace{-0.3 cm}
\renewcommand{\arraystretch}{1.5} 
\begin{subtable}[h]{\textwidth}
\begin{tabular}{l  l | l  l | l l}
    \hline
    \hline	
    $J_{1}$                  & 21        & $J_{2}$                  &    84                 & $\bm{p}_{\rm{BS}}$             & [0,0, 2]             \\
    $Q$                      & 12        & $W_{\rm{o}}$             & 15 KHz                & $W_{\rm{sub}}$                 & 750 KHz              \\
    $N_{\rm{BS}}^{(\rm{x})}$ & 24        & $N_{\rm{BS}}^{(\rm{z})}$ & 24                    & $z_{\rm{UE}}$                  & $\in [1,1.5]$        \\
    BW                       & 8.265 MHz & $\delta_{\rm{BS}}$       & $\lambda_{\rm{o}} /2$ & $\theta_{\rm{UE}}^{(\rm{az})}$ & $\in [-\pi/4,\pi/4]$ \\
    $f_{\rm{o}}$             & 24 GHz    & $\rm{Noise}_{\rm{F}}$    & 10 dB                 & $\rm{Noise}_{\rm{D}}$          & -174 dBm/Hz          \\
    \hline
\end{tabular}
\end{subtable}
\end{center}
\end{table}

\begin{figure}[h]
    \centering
    \includegraphics[width=0.9\columnwidth]{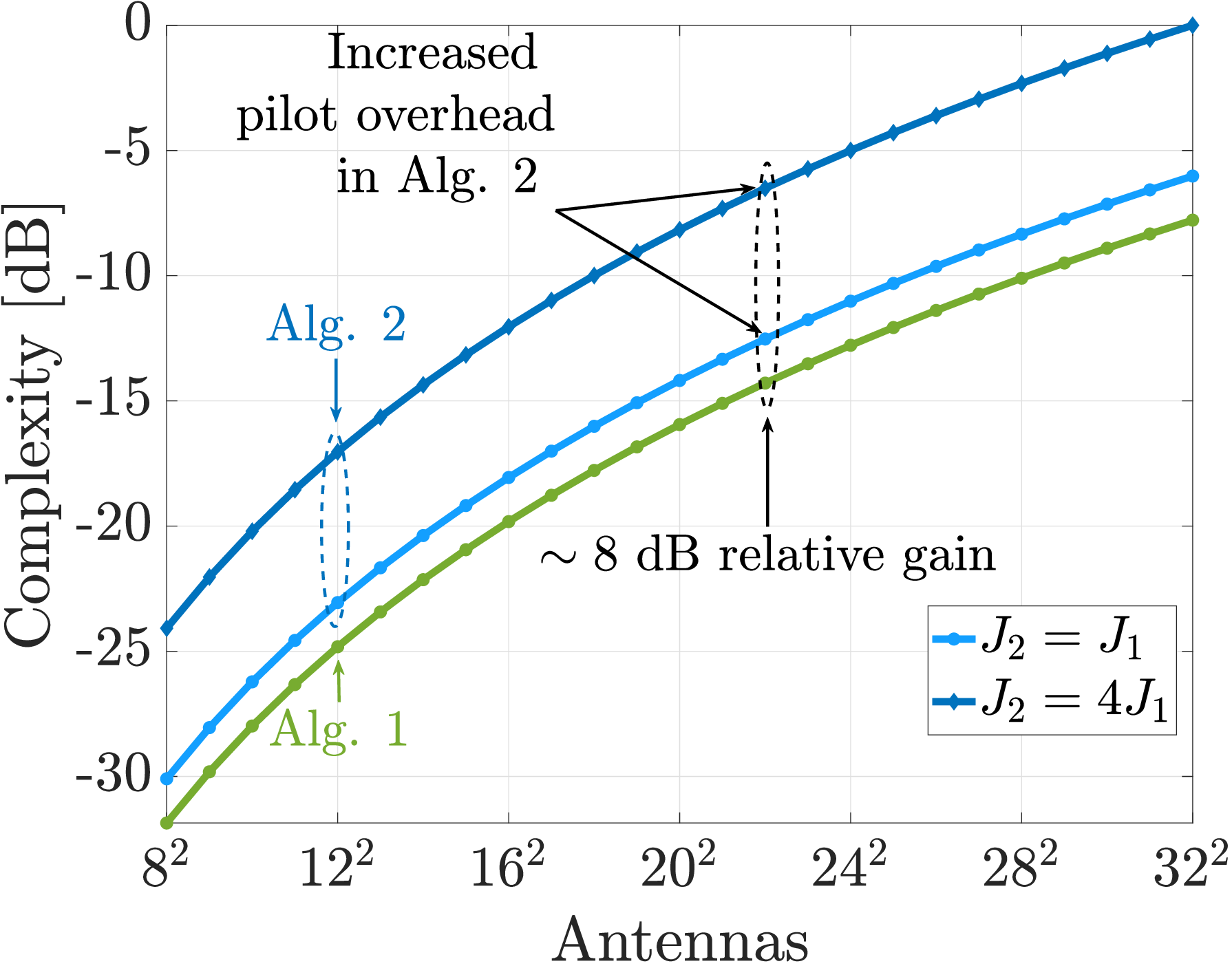}
    \caption{Normalised relative complexity for different signaling overhead requirements for the two proposed Algorithms.}
    \label{fig:comp}
\end{figure}

Figure~\ref{fig:comp} shows the computational advantage of the FF-based localization algorithm. The complexity rises with larger BS arrays and higher signaling overhead requirements. 
In our simulation set-up, the FF-based scheme has 75\% lower signaling overhead, which results in, approximately 8 dB of computational gain, i.e., when $J_{2}=4J_{1}$ the run time of Alg.~\ref{alg:pos_scheme_2} is 6.3 times higher than Alg.~\ref{alg:pos_scheme_1}. A computational gain is observed for similar signaling overheads as well. The values here are normalized relative to the largest value. Overall, a definitive closed-form expression can not be derived and it needs to be noted that both algorithms do not produce optimal estimations, but rather they are designed with a similar practical structure that highlights the limitation of beam-steering generalizations and relaxed channel assumptions in localization applications. 

\begin{figure}[h]
    \centering
    \includegraphics[width=0.9\columnwidth]{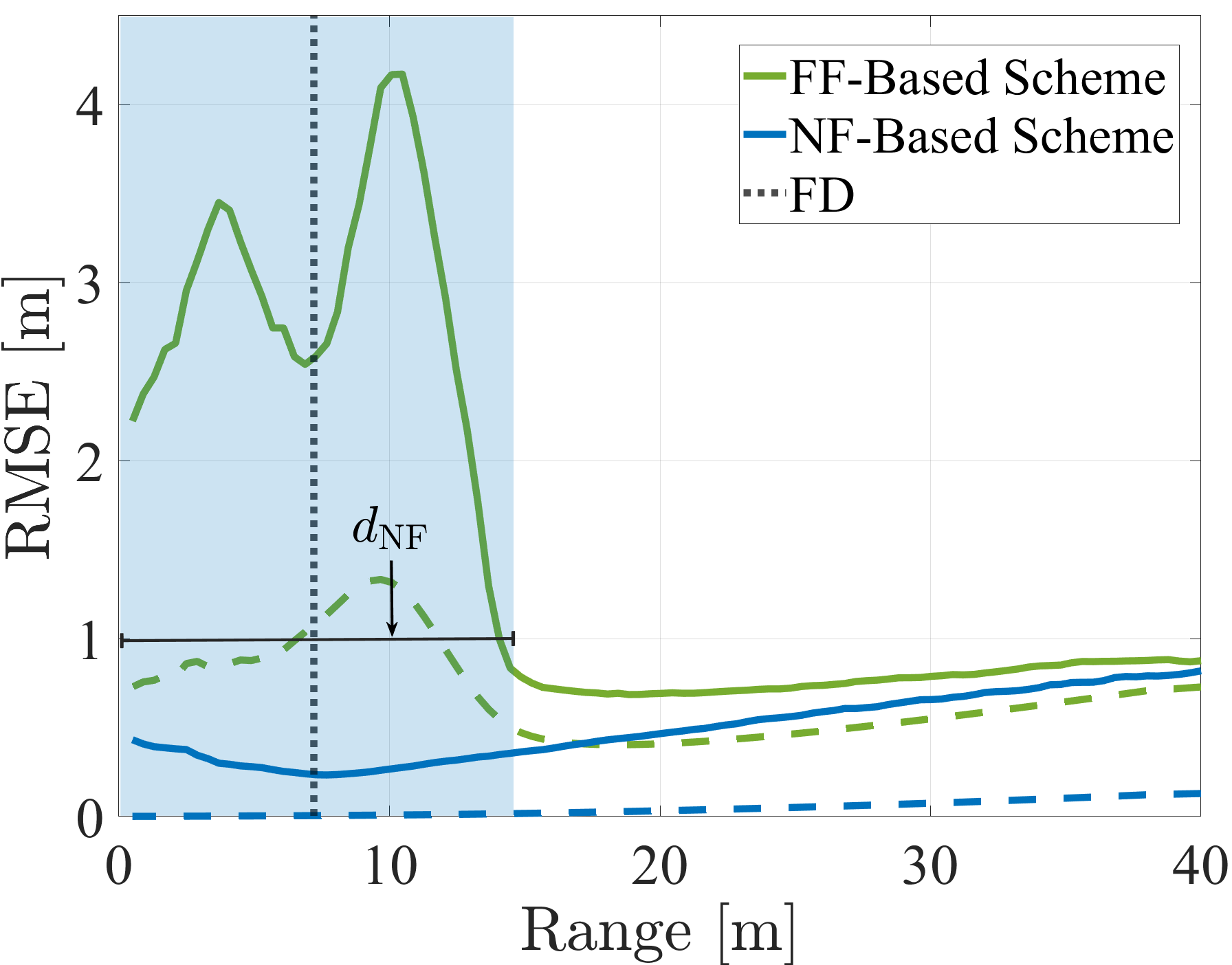}
    \caption{Position RMSE for the two proposed signaling schemes. The dashed lines mark the corresponding PEBs.}
    \label{fig:pos_rmse}
\end{figure}

Figure~\ref{fig:pos_rmse} shows the positioning root mean square error (RMSE) in meters, for the two signaling schemes, versus the BS-to-UE distance. Since more resources are utilized in the NF-based scheme, it consistently outperforms the FF-based scheme. However, the performance of the two proposed schemes notably converge and the performance gain of the NF-based scheme decreases as the BS-to-UE distance increases. The FD distance is marked to highlight that it is not a good metric for the compatibility of the FF-based signaling design and the corresponding positioning algorithm. In a practical scenario, the NF-based scheme should be employed when the expected performance gain out-weights the additional signaling requirements and increased computational load. In Fig.~\ref{fig:pos_rmse}, the shaded region marks the range that the NF-based scheme \emph{needs to be employed}, for such an approach. It is shown that both schemes could be employed, depending on the desired accuracy and the UE's position relative to the BS. For comparison, the position error bound (PEB) is shown when the true channel model in~\eqref{eq:channel_h_NF} and the relaxed channel model in~\eqref{eq:channel_h_FF} are used. Since the structure of the two algorithms is similar and the additional signaling of the NF-based scheme is focused close to the BS, their performance converges as the distance increases.

\begin{figure}[h]
    \centering
    \captionsetup{justification=centering}
     \centering
     \includegraphics[width=0.85\columnwidth]{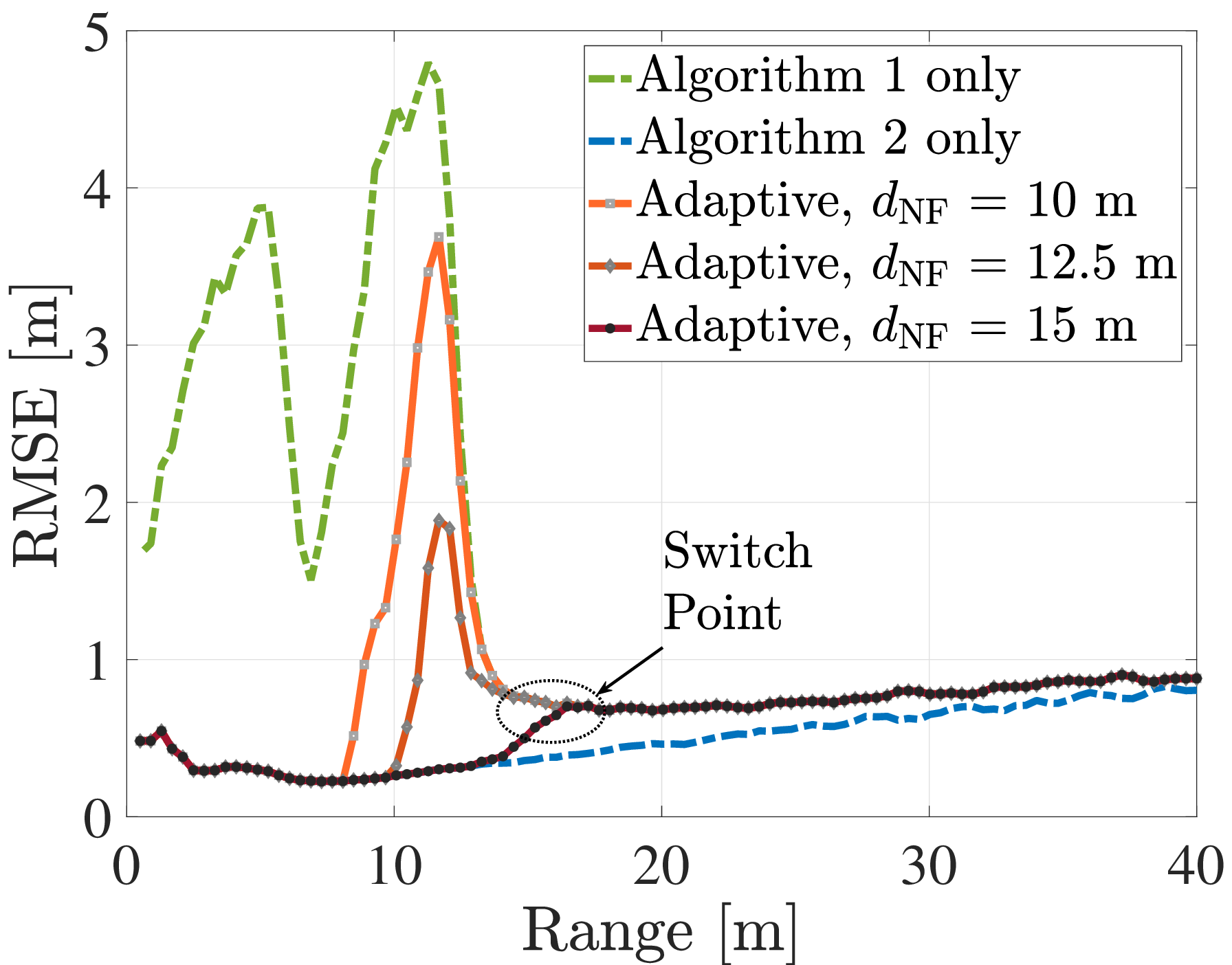}
     \caption{RMSE for the proposed adaptive procedure, as the UE is moving towards the BS.}
     \label{fig:track_rmse}
\end{figure}

Figure~\ref{fig:track_rmse} shows the RMSE for an iterative scenario, where a UE moves within $\mathcal{R}$. The compatibility of the FF-based scheme is compromised when the UE is close to the BS and there is a performance gap between the two schemes. Using exclusively the FF-based scheme (resp. NF-based scheme) marks the limit in lowest (resp. highest) accuracy and complexity for our proposed adaptive scheme. Our adaptive procedure avoids to employ the FF-based scheme when it is not reliable. The UE starts in the FF and maintains the FF-based scheme as it moves towards the BS. The size of the NF region, $d_{\rm{NF}}$ varies. Increasing $d_{\rm{NF}}$ improves the system's performance at the expense of more resources, as the NF-based scheme is employed earlier. If the switching point is placed closer to the FF-NF region the low complexity FF-based scheme is utilized more, but there is an increased probability that unreliable estimations will be also used. On average the FF-based scheme was employed for $25.82\%$, $31\%$ and $37.32\%$ of the procedure when $d_{\rm{NF}}$ is set to 10 m, 12.5m and 15 m, respectively. This highlights the accuracy-to-resources trade off and the degrees of freedom in the design of the adaptive procedure. The effectiveness of this iterative procedure could be improved if a sense of memory was introduced and prior information was jointly processed.

\begin{figure}[h]
    \centering
    \includegraphics[width=0.85\columnwidth]{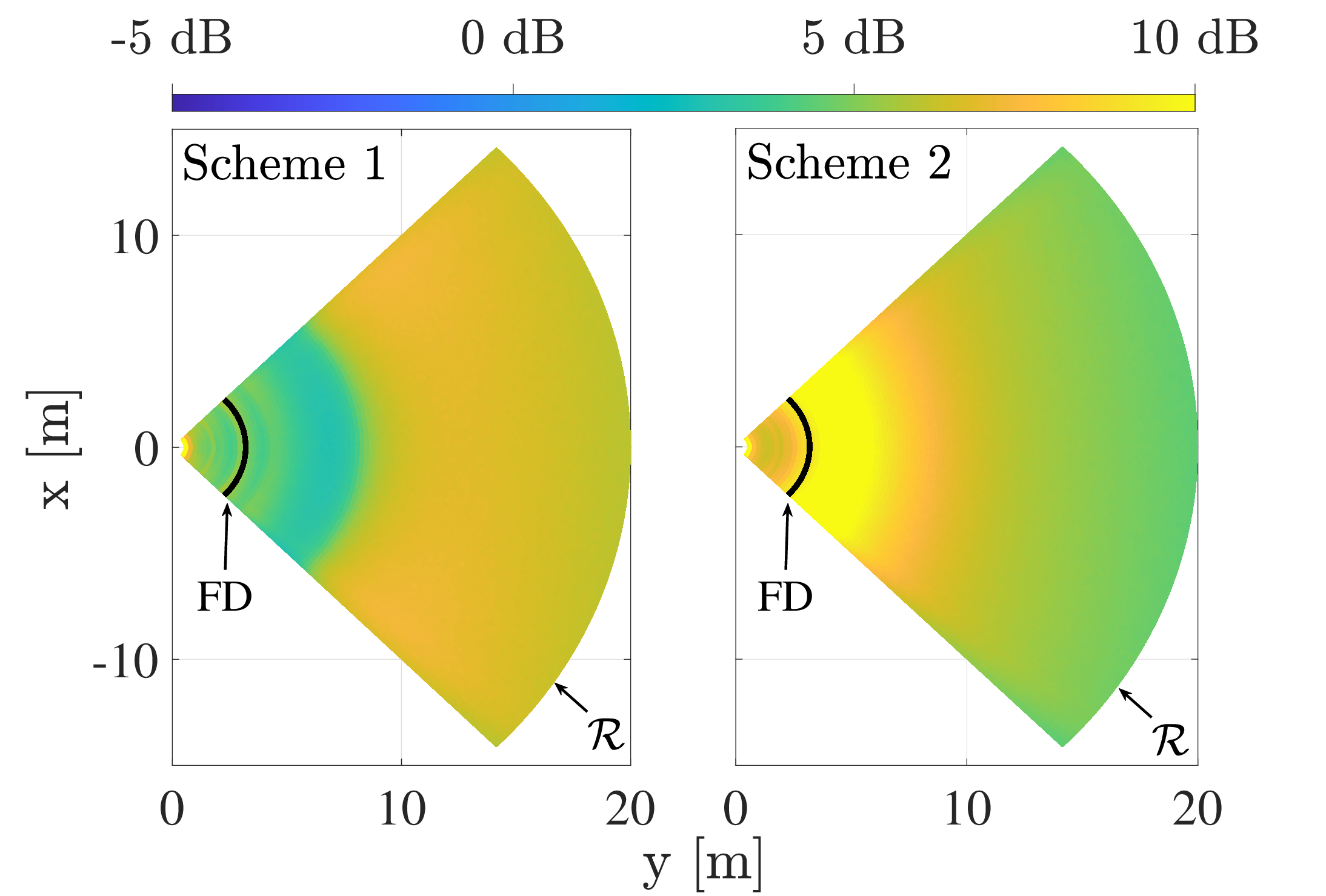}
    \caption{SNR for the two different signaling schemes.}
    \label{fig:snr}
\end{figure}

Figure~\ref{fig:snr} shows the signal to noise ratio (SNR) for the two proposed signaling schemes. It is shown that the FF-based beam design is problematic for small the BS-to-UE distances. Despite the smaller path-loss, UEs close to the BS suffer from low SNR. The NF-based scheme appears to successfully combat this issue, as more beams and sufficiently dense focus points are considered. The beam-steering technique of the FF-based scheme and the beam-focusing technique of the NF-based scheme display similar behaviour for larger distances, as highlighted by the similar SNR values deeper in $\mathcal{R}$. The structure of $\mathcal{R}$ affects the design process of the two beam-sweeping strategies and the FD is shown to be an unreliable indicator for the compatibility of the beam-steering design.
\section{Conclusions}
In this paper, we proposed two DL localization schemes with different computational complexities and signaling overheads. Each scheme is distinctive  in terms of FF/NF conditions and the BS's ability to effectively employ a beam-steering rather than a beam-focusing technique. We investigated the trade-off between complexity and accuracy highlighted that the conventional definition of the FF region fails to dictate a practical localization scheme that is based on the BS-to-UE distance. Hence, the FD distance is not necessarily a good metric for the validity of the FF/NF assumption. In addition, we explored an iterative adaptive procedure utilizing the appropriate signaling scheme. The higher complexity scheme is employed when the expected performance gain justifies the additional resource requirements and the unreliable region of the FF-based scheme is effectively avoided. 
\section*{Acknowledgement}
This work has received funding from the European Union’s Horizon 2020 research and innovation
programme under the Marie Skłodowska-Curie grant agreement No 956256

\bibliographystyle{ieeetr}
\bibliography{references}

\end{document}